\documentclass[12pt, preprint, onecolumn]{emulateapj}
\usepackage{natbib}
\usepackage{graphicx}
\usepackage{amsmath}
\usepackage{xcolor}
\usepackage{url} 
\usepackage{wrapfig}
\usepackage{placeins}
\usepackage{longtable}
\usepackage[caption = false]{subfig}
\usepackage[colorlinks = true, linkcolor = blue, urlcolor = blue]{hyperref}

\def\gtaprx{ \mathrel{ \vcenter{
      \offinterlineskip \hbox{$>$}
      \kern 0.3ex \hbox{$\sim$}    } } }

\def\ltaprx{ \mathrel{ \vcenter{
      \offinterlineskip \hbox{$<$}
      \kern 0.3ex \hbox{$\sim$}    } } }

\shorttitle{Hypatia Catalog Database}
\shortauthors{Hinkel \& Burger}

\begin{document}
\title{The Hypatia Catalog Database: \\ A Web-Based Interface for Exploring Stellar Abundances}

\author{Natalie R. Hinkel\altaffilmark{1} \&
Dan Burger\altaffilmark{1}}

\altaffiltext{1}{Physics \& Astronomy Dept,
                 Vanderbilt University,
                 Nashville, TN 37235, USA}
\email{For questions or comments: hypatiacatalog@gmail.com}

\section{Introduction}
\label{s.intro}
\begin{figure}[h]
\begin{center}
\vspace{5mm}
 \centerline{\includegraphics[width=17cm]{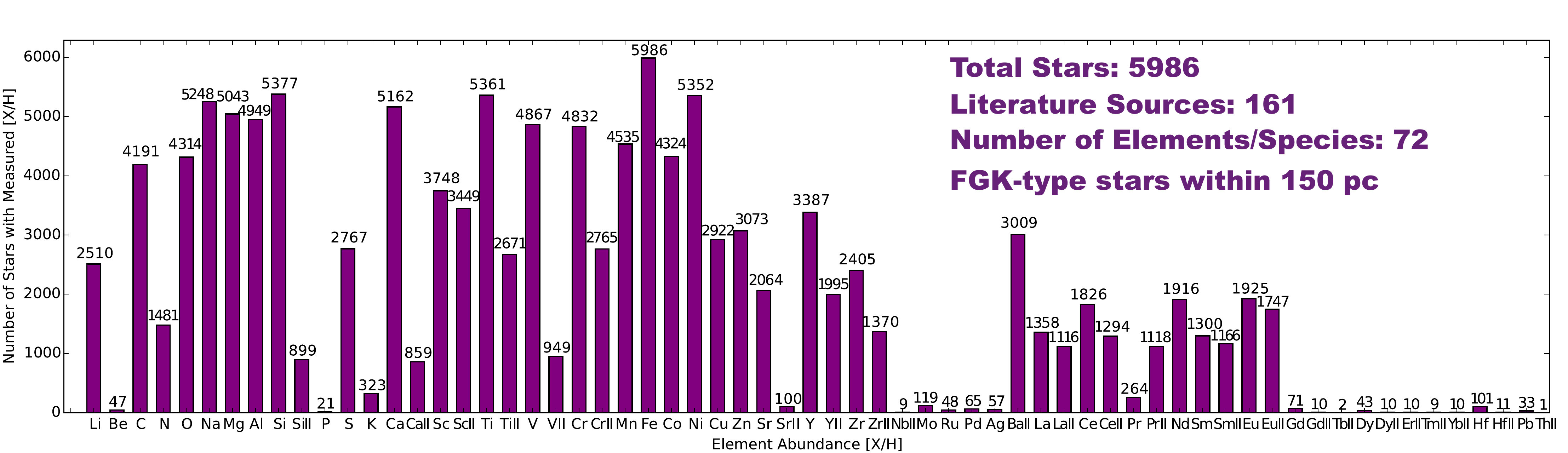}}
\end{center}
  \caption{Histogram showing the current elements and the total number of stars in which they have been measured within the Hypatia Catalog and the Hypatia Catalog Database.} \vspace{3mm}
  \label{hist}
\end{figure}

The Hypatia Catalog Database (\url{www.hypatiacatalog.com}) is the largest database of high resolution stellar abundances for stars within the solar neighborhood. It currently offers 72 elements and species within 5,986 stars, 347 of which host planets, as shown in Fig. \ref{hist}. The Hypatia Catalog Database features an interactive table and multiple plotting interfaces that allow easy access and exploration of stellar abundance data and properties within the Hypatia Catalog. The Hypatia Catalog is a multidimensional, amalgamate dataset comprised of stellar abundance measurements for FGKM-type stars within 150 pc of the Sun from carefully culled literature sources, currently totaling 161 datasets, that measured both [Fe/H] and at least one other element. In addition to abundances, stellar properties and planetary properties, where applicable, have been made available within the Hypatia Catalog Database. Data can be downloaded currently through the website for use in personal plotting routines, to be made accessible through the terminal in the near future. 


A detailed description of the Hypatia Catalog can be found in \citet{Hinkel14} and was subsequently updated in \citet{Hinkel16, Hinkel17a}. The Hypatia Catalog and Hypatia Catalog Database will continue to be routinely updated in order to incorporate the most recent stellar abundance data published within the literature. Help and documentation for the Hypatia Catalog Database can be found in Section \ref{s.plots} or under the ``Help" link in the top right corner of the website. A more detailed discussion of the data is in Section \ref{s.about} or on the ``About" page of the website. General appreciation and acknowledgments to be included in published papers can be found in Section \ref{s.ack} or under the ``Acknowledgments" link. Finally, for any website or data updates, issues, or corrections, please email \url{hypatiacatalog@gmail.com}.

\section{Plots and Table}
\label{s.plots}
The Hypatia Catalog Database offers a seamless, interactive website in order to view stellar abundance data both graphically and in a tabular format. Below we discuss the plotting interfaces and data table. 

\subsection{Elements \& Properties}
\begin{figure}[h]
\begin{center}
 \centerline{\includegraphics[width=17cm]{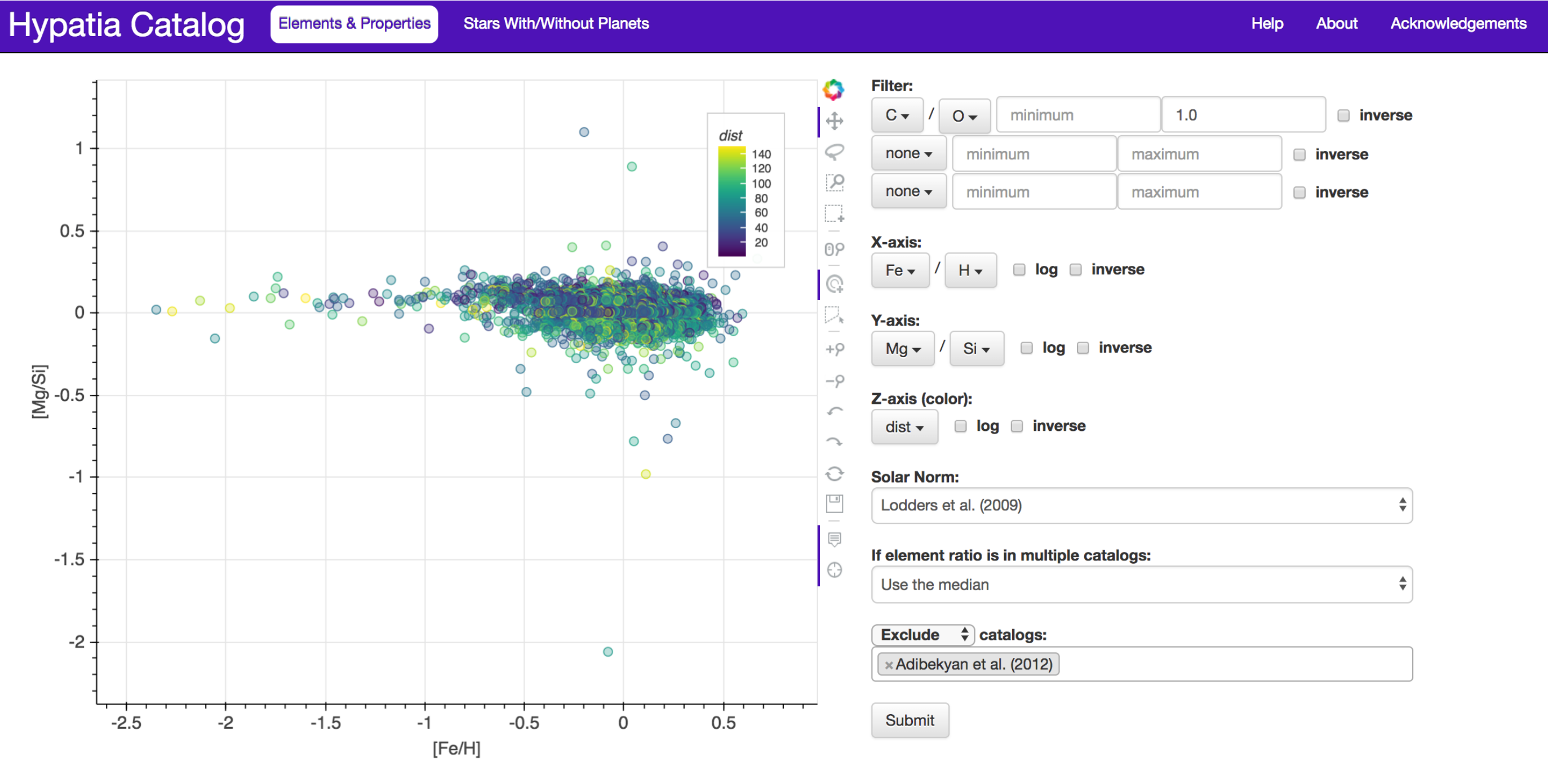}}
\end{center}
  \caption{Example plot on the ``Elements \& Properties" tab showing how to use the filter, vary the X- and Y-axes, include a color bar for distance on the Z-axis, and exclude a particular dataset$^{2}$.} \vspace{3mm}
  \label{plot1}
\end{figure}

To begin, the plot can be filtered using the 3 filter boxes on the top right that provide minimum and maximum values (either can be left blank in place of infinity). The ``Submit" button must be clicked in order to enact changes. Options for the filter criteria can be accessed by clicking on the dropdown menu that says ``none" as default. From the dropdown menu, if an element is chosen from the Periodic Table, the options to filter changes to be an X/Y ratio. In addition to the neutral state of the element, if the singly ionized element has been measured, the ``II" next to the element can be selected. Stellar properties are listed along the bottom of the drop down menu in addition to planetary properties. If any of planetary properties are chosen, then the graph will show only planet hosting stars. The X and Y axes can plot any of the element ratios, stellar properties, or planet properties. Additionally, the Z-axis, or color, can be initialized to show another dimension. Finally, while the Hypatia Catalog and Database is unbiased in its inclusion of datasets, we recognize that the user may wish to look exclusively at particular catalogs. We have provided a toggle that allows the user to switch between ``Excluding" or ``Allowing" individual catalogs as they see fit. See Figure \ref{plot1} for a demonstrated plot using the filter, the color bar, and the exclude option for the \citet{Adibekyan12} catalog\footnote{Our apologies to Vardan Adibekyan et al. (2012) for using their dataset in this demonstration. This in no way reflects our esteem for their paper or data (which is high!), but was merely chosen because we thought he/they would forgive us at some point in the near future.}.

To move, pan, zoom, or generally change the properties of the plot, the tools listed vertically along the right side of the plot can be toggled to make the appropriate adjustments. Additionally, the figure can be saved by clicking the ``Save" button near the bottom. 

\subsection{Stars With \& Without Planets}
\begin{figure}[h]
\begin{center}
 \centerline{\includegraphics[width=17cm]{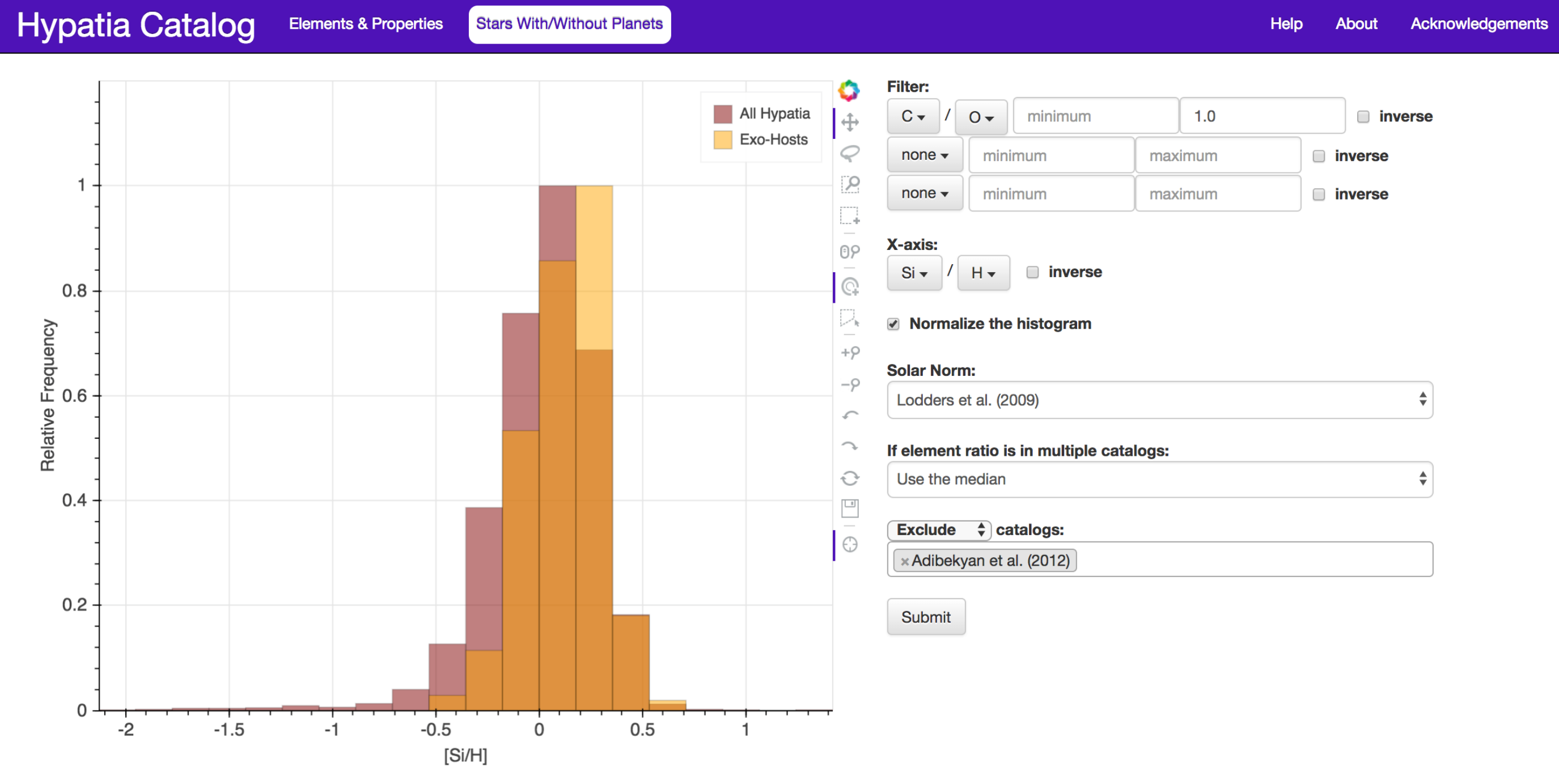}}
\end{center}
  \caption{The 1D histogram on the ``Stars With \& Without Planets" tab.} \vspace{3mm}
  \label{plot2}
\end{figure}

While the plot showing ``Elements \& Properties" enables a user to look at either planet hosting stars or stars that are not currently known to have planets, we wanted to provide the ability to look at the two populations simultaneously. Therefore, we have provided a 1D histogram that shows both stars with and without planets, shown in Fig. \ref{plot2}. Similar to the other plot, the filter bar can be used to make cuts on those stars analyzed. Additionally, the histogram can viewed either in terms of total number of stars in a bin or normalized (by checking the ``Normalize the histogram" toggle) according to the bin with most number of stars, which has a height of 1 by definition.  Note that all decisions and filters used for the ``Elements \& Properties" plot are ported to the ``Stars With \& Without Planets" tab. 

\subsection{Advanced Controls}
For those users wishing to change some of the more nuanced aspects of the stellar abundances within the Hypatia Catalog and Database, we provide the ability to make adjustments. In order to ensure that all of the stellar abundances within Hypatia are on the same baseline, all individual catalogs have been renormalized to the same solar normalization. The default is \citet{Lodders:2009p3091}. As part of the database, the user can change the solar normalization to be with respect to \citet{Asplund:2009p3251, Asplund:2005p7415, Grevesse:1998p3102}, and \citet{Anders:1989p3165}. If the user would like to view the stellar abundances with respect to the solar normalization employed by each of the catalogs individually, the ``Internal" option can be chosen. Similarly, the absolute abundances, namely without a solar normalization, can be viewed by choosing ``Absolute."

The Hypatia Catalog is a multidimensional database, which utilizes all stellar abundances measurements made for an element within a star by +200 catalogs \citep[see][, Fig 3]{Hinkel14}. Therefore, to provide a 2-dimensional table, those measurements must be collapsed or reduced. The toggle ``If an element ratio is in multiple catalogs" changes the reduction method to either take the median (default) or mean of all measurements of an element in a star.

\subsection{Table}
\begin{figure}[h]
\begin{center}
 \centerline{\includegraphics[width=17cm]{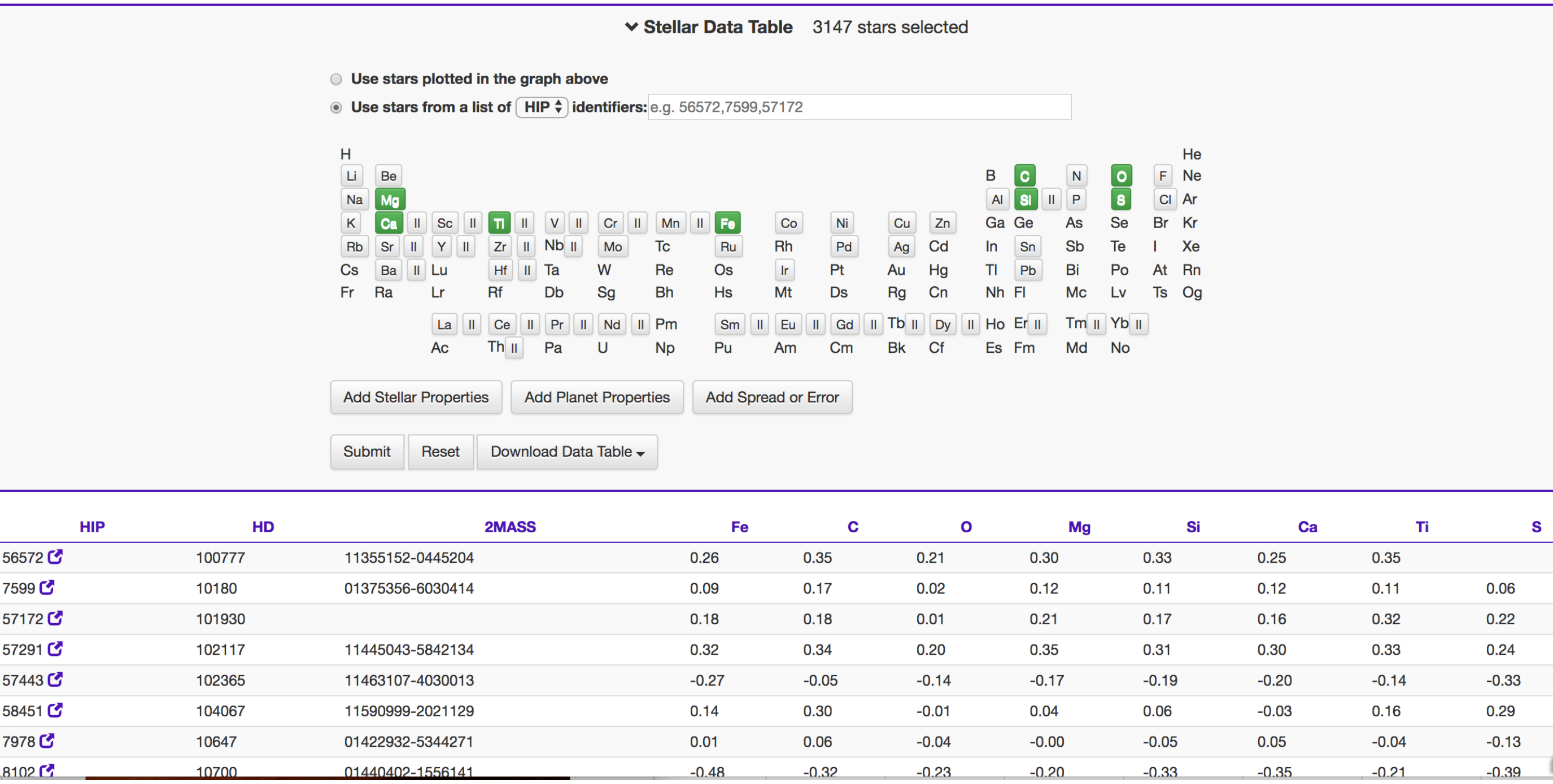}}
\end{center}
  \caption{Screenshot showing the column options for the data table. Filtering of the table can be done through the filter bars next to the ``Elements \& Properties" plot, which is above the table.}  \vspace{3mm}
  \label{table1}
\end{figure}

A table of the data, see Figure \ref{table1}, lies just below the plot on the website and can filtered using the filter bars at the top. The columns that are shown in the table can be adjusted using the ``Add/remove columns" button, which gives the user the ability to include individual elements (all [X/H] unless the Absolute solar normalization is chosen, see below). Additionally, all stellar properties (RA, Dec, XYZ position, distance, disk, spectral type, V mag, B-V, UVW galactic velocity, Teff, and logg) and all planetary properties (planet letter, period, planet mass, eccentricity, and semimajor axis, where applicable) can be given in the table. Finally, the user can also view the spread (or error) in the stellar abundances. Per the Hypatia Catalog \citep{Hinkel14}, the spread is defined as the range in the stellar abundance measurements when determined by different groups measuring that same element within that same star. The spread is an exceptionally useful tool to truly define how well an abundance measurement is ``understood" within the star \citep[see][for more details]{Hinkel16}.

\section{About the Data and Hypatia}\label{s.about}
In Table \ref{params} we've listed the column labels used in the Hypatia Catalog Database, the associated units, and a description of the data with source, where applicable.  For a complete breakdown of all catalogs (with $>$ 20 stars) included in the Hypatia Catalog and Hypatia Catalog Database, please \href{https://drive.google.com/file/d/0B6JubEpe8-76YjlBakJydy1OaXM/view}{download this file}, the components of which can be found in \citet{Hinkel14,Hinkel16,Hinkel17a}. In the header, ``S/N" is signal-to-noise report by the literature source, ``$\lambda$ Range" is the wavelength coverage, "Stellar Atmo" is the stellar atmospheric model, "Eq. Width" is the package used to determine the equivalent width, "CoG or SF" designates whether the group used a curve-of-growth or spectral fitting technique where the package is specified in the former case, the "Solar Scale" is the solar normalization used by that group (differential analysis is cited where applicable), and "Num. of Fe I/II lines" lists the number of Fe I and Fe II lines. While this accumulation of telescope information, reduction packages, and overall technique was done with the best intentions, errors may arise so please check the original source for verification. Please email hypatiacatalog@gmail.com with any questions or concerns. 

Per the ionization states, while some catalogs measured only one ionization state when reporting an abundance determination, a number of catalogs combined the abundances from multiple ionization states. In Hypatia, an abundance of [X/Fe] means that a catalog measured the neutral state, a combination of neutral and ionized state(s), or it was not specified. Whenever a catalog specifically mentioned it was only measuring the singly ionized state, we demarcate it as [X II/Fe]. 

A note about binaries, in the event that abundances were measured within a stellar binary system (i.e., labeled ``A" and ``B" in the naming scheme), the abundances for the ``A" component were preferentially chosen while the ``B" component was ignored. This was done for computational reasons, not scientific, and so this feature may be updated in a later iteration.

Finally, the Hypatia Catalog and Hypatia Catalog Database were named for Hypatia of Alexandria, one of the first known female astronomers. She was strong-minded, independent, and well respected by both the people and local officials. Socrates Scholasticus\footnote{Socrates Scholasticus is not to be confused with Socrates, the Greek philosopher who lived 800 years before Hypatia's time.} once said of her, ``For all men on account of her extraordinary dignity and virtue admired her the more."  For more information, please \href{http://www.nataliehinkel.com/hypatia-catalog.html}{see this page}, where more resources can be found.

\section{Acknowledging the Hypatia Catalog Database}\label{s.ack}
If you have used the Hypatia Catalog Database in published research, please cite the original Hypatia Catalog paper per \citet{Hinkel14} where appropriate and include the following acknowledgement: \\
``The research shown here acknowledges use of the Hypatia Catalog Database, an online compilation of stellar abundance data as described in Hinkel et al. (2014, AJ, 148, 54), which was supported by NASA's Nexus for Exoplanet System Science (NExSS) research coordination network and the Vanderbilt Initiative in Data-Intensive Astrophysics (VIDA)."


\begin{wrapfigure}{r}{0.4\textwidth}
\begin{center}
 \vspace{-7mm}
 \centerline{\includegraphics[width=0.3\textwidth]{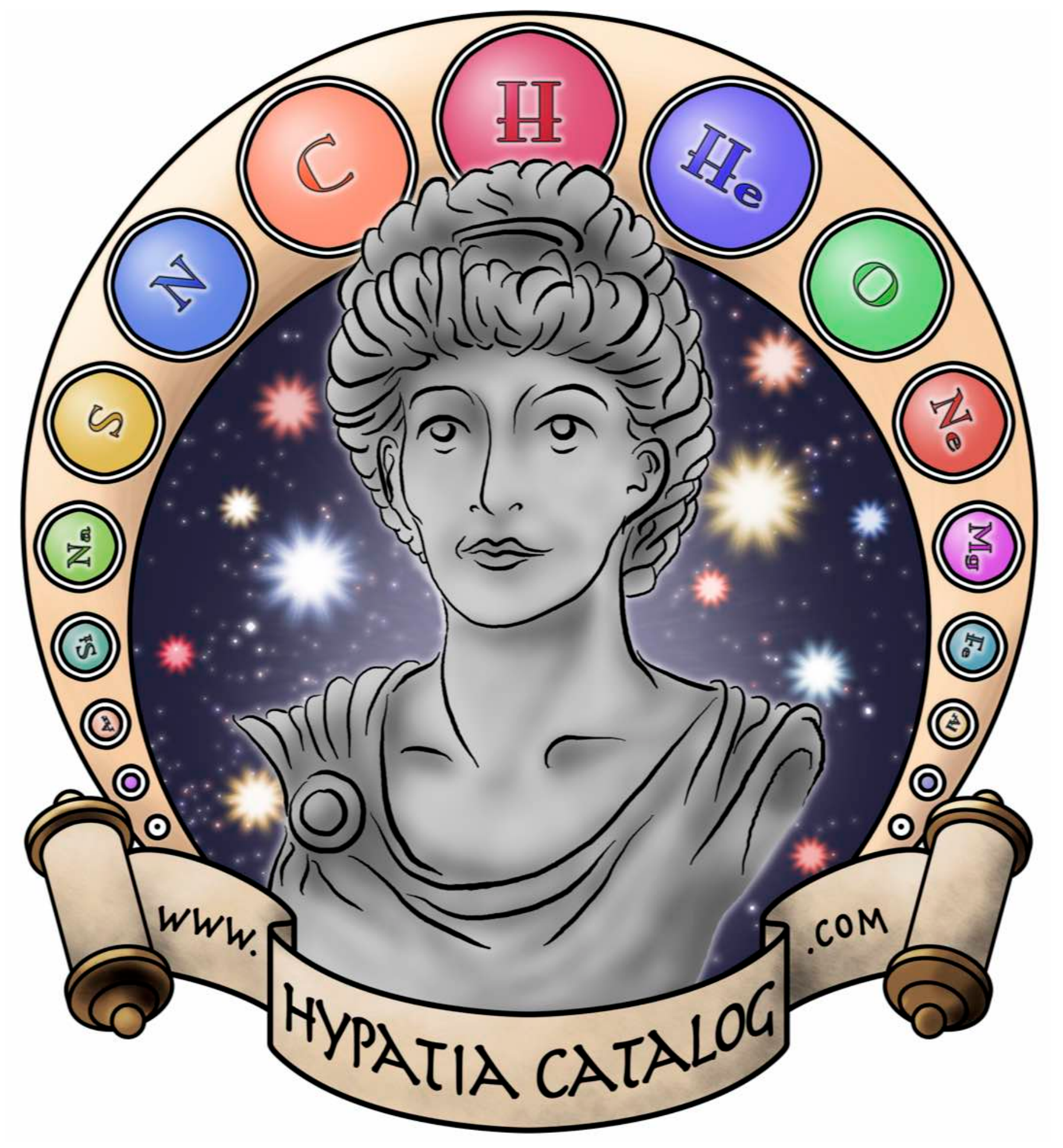}}
  \vspace{-5mm}
\end{center}
  \caption{Hypatia Catalog Database logo made by Nahks Tr'Ehnl.}  \vspace{3mm}
  \label{logo}
\end{wrapfigure}
In the event that a small number of catalogs was used within an analysis, we highly encourage the citation of those individual papers. However, we still appreciate the acknowledgment of the Hypatia Catalog and database, in order to give credit for the use of the website, internal algorithms, and availability of the data (since +50\% of the catalogs were not available in machine-readable formats prior to this compilation).

The Hypatia Catalog and Hypatia Catalog Database was created and is currently maintained by Natalie Hinkel. The lead developer for the online Hypatia Catalog Database is Dan Burger, who acknowledges support from the Vanderbilt Initiative for Autism, Innovation and the Workforce.  Additional developer support by Nitin Pasumarthy, Benjamin Knight, and Wendy Yu. Logo by Nahks TrÕEhnl (see Fig. \ref{logo}), who can be found at \url{www.nahks.com}. Additional thanks to Kevin White, Garrett Somers, and Caleb Wheeler.  Please send any website or data updates, issues, or corrections to hypatiacatalog@gmail.com. You can also find us on Twitter via @hypatiacatalog.

\newpage
\LongTables
\begin{deluxetable*}{p{1.4cm}p{1.4cm}p{10.0cm}}
  \tablecaption{\label{params} Parameters in the online Hypatia Catalog Database}
  \tablehead{
  \colhead{Label} &
    \colhead{Unit} &
    \colhead{Description}  
  }
 \startdata
HIP & --- & Hipparcos name \\
HD & --- & Henry-Draper catalog name \\
2MASS & --- & 2MASS name \\
RAdeg & deg & Right Ascension decimal degrees (J2000) \\
DEdeg & deg & Declination decimal degrees (J2000) \\
X & pc & Cartesian geocentric x-coordinate from the Sun \\
Y & pc & Cartesian geocentric y-coordinate from the Sun \\
Z & pc & Cartesian geocentric z-coordinate from the Sun \\
Dist & pc & distance in pc (from Gaia and \citealt{Anderson12}) \\
pmRA & km/s & proper motion for Right Ascension (from Gaia per \citealt{Arenou17} and \citealt{Anderson12}) \\
pmDec & km/s & proper motion for Declination (from Gaia per \citealt{Arenou17} and \citealt{Anderson12}) \\
UVel & km/s & Component of space velocity positive toward the Galactic anticenter radial (from \citealt{Anderson12}, null = 9999.0) \\
VVel & km/s & Component of space velocity positive in the direction of Galactic rotation (from \citealt{Anderson12},  null = 9999.0) \\
WVel & km/s & Component of space velocity positive toward the North Galactic Pole (from \citealt{Anderson12},  null = 9999.0) \\
Teff & K & stellar effective temperature (from PASTEL per \citealt{Soubiran16}) \\
logg & cm/s$^2$ & surface gravity of the star (from PASTEL per \citealt{Soubiran16}) \\
Disk & --- & likely origin within the disk/thin/thick based on kinematics (null = N/A) \\
SpT & --- & spectral type (from \citealt{Anderson12}) \\
Bmag & mag & B magnitude (from PASTEL per \citealt{Soubiran16}) \\
Vmag & mag & V magnitude (from PASTEL per \citealt{Soubiran16}) \\
B-V & mag & B-V color (from PASTEL per \citealt{Soubiran16}) \\
letter & --- & planet letter (from NASA Exoplanet Archive) \\
multi & --- & number of planets in the system (from NASA Exoplanet Archive) \\
pmass & M$_{Jupiter}$ & planetary mass (from NASA Exoplanet Archive) \\
period & days & planetary period (from NASA Exoplanet Archive) \\
ecc & --- & planetary eccentricity (from NASA Exoplanet Archive) \\
sma & AU & planetary semimajor axis (from NASA Exoplanet Archive) \\
smass & M$_{\odot}$ & stellar mass (from NASA Exoplanet Archive) \\
radius & R$_{\odot}$ & stellar radius (from NASA Exoplanet Archive) \\
XH & dex & [X/H] abundance in dex as compiled from multiple catalogs (which can be included/excluded in plots) \\
e\_XH & dex & Spread in XH in dex or the range in abundance measurements as reported by multiple catalogs
 \enddata
 \end{deluxetable*} \vspace{3mm}

\providecommand{\noopsort}[1]{}

\end{document}